\documentclass[superscriptaddress,secnumarabic,amssymb,amsmath,nobibnotes,aps,prd,showkeys,nofootinbib,preprint]{revtex4}
\usepackage{graphicx}
\usepackage{float}
\usepackage{bm}
\usepackage{amsmath}
\usepackage{amsfonts}
\usepackage{amssymb}
\usepackage{epstopdf}
\usepackage{natbib}%
\newcommand{\bee}{\begin{equation}}
\newcommand{\eee}{\end{equation}}
\newcommand{\eaa}{\end{eqnarray}}
\newcommand{\baa}{\begin{eqnarray}}
\def\ni{\noindent}

\usepackage{color}

\begin{document}
	

\title{\large Statistical approaches on the apparent horizon entropy and the generalized second law of thermodynamics}

\author{Everton M. C. Abreu}\email{evertonabreu@ufrrj.br}
\affiliation{Departamento de F\'{i}sica, Universidade Federal Rural do Rio de Janeiro, 23890-971, Serop\'edica, RJ, Brazil}
\affiliation{Departamento de F\'{i}sica, Universidade Federal de Juiz de Fora, 36036-330, Juiz de Fora, MG, Brazil}
\affiliation{Programa de P\'os-Gradua\c{c}\~ao Interdisciplinar em F\'isica Aplicada, Instituto de F\'{i}sica, Universidade Federal do Rio de Janeiro, 21941-972, Rio de Janeiro, RJ, Brazil}
\author{Jorge Ananias Neto}\email{jorge@fisica.ufjf.br}
\affiliation{Departamento de F\'{i}sica, Universidade Federal de Juiz de Fora, 36036-330, Juiz de Fora, MG, Brazil}

\keywords{Tsallis statistics, R\'enyi statistics, Kaniadakis statistics, Sharma-Mittal statistics, generalized second law}

\begin{abstract}
\noindent In this work we have investigated the effects of three nongaussian entropies, namely, the modified R\'enyi entropy (MRE), the Sharma-Mittal entropy (SME) and the dual Kaniadakis entropy (DKE) in the investigation of the generalized second law (GSL) of thermodynamics violation. The GSL is an extension of the second law for black holes. 
Recently, it was concluded that a total entropy is the sum of the entropy enclosed by the apparent horizon plus the entropy of the horizon itself when the apparent horizon is described by the Barrow entropy. It was assumed that the universe is filled with matter and dark energy fluids. Here, the apparent horizon will be described by MRE, SME, and then by DKE proposals.  Since GSL holds for usual entropy, but it is conditionally violated in the extended entropies, this implies that the parameter of these entropies should be constrained in small values in order for the GSL to be satisfied.
Hence, we have established conditions where the second law of thermodynamics can or cannot be obeyed considering these three statistical concepts just as it was made in Barrow's entropy. 
Considering the $\Lambda CDM$ cosmology we can observe that for MRE, SME and DKE, the GSL of thermodynamics is not obeyed for small redshift values.
\end{abstract}

\maketitle

\section{Introduction}

One of the most pivotal discoveries of the last decades is that the universe is accelerating \cite{perlmutter}.   Considering thermal properties of an accelerated universe \cite{sadjadi}, the investigation of the generalized second law (GSL) defining black holes (BHs) thermodynamics has increased \cite{dps}.   Namely, there exist theoretical proofs concerning the thermodynamical features of BH.   These connections are represented by both the specific temperature and entropy computation by using the BH horizon \cite{gh}.   The Friedmann equations can be obtained through the first law of thermodynamics and the apparent horizon.   On the other hand, we can write the Friedmann equations as the first law \cite{friedmann}, which worked in general relativity (GR) and its alternative propositions. 

The GSL of thermodynamics was, at the beginning, depicted for BHs \cite{bek,jdb}.   The GSL established that the sum of the standard entropy plus one quarter of the area (A) of the event horizon cannot decrease with time where we can use the natural units, namely, $8\pi G=\hbar=c=k=1$.   Hence, $2\pi A$ is connected to the gravitational entropy of the BH.   In \cite{gh} the authors assumed that we can connect a gravitational entropy also to de Sitter space.   And the same expression $2\pi A$ was used, where now $A$ is the area of the de Sitter horizon.   The study of the swap of entropy and energy between heat baths and BHs or de Sitter horizon make sense in the scenarios of these ideas \cite{gh,hawking,davies}.

Statistical GSL violations are not impossible.   However, they turn out to be highly improbable since the systems becomes larger and larger.   Hence, GSL is a statistical law where violations can happen due to large fluctuations.   The violations of GSL means that it is not guaranteed a positivity for entropy, namely, it means that the total entropy can be not a non-decreasing function of time.   So, since GSL of thermodynamics is underlying in physics, its violation means a strong factor against the fundamental theory.
 
Considering the GSL of thermodynamics by Bekenstein \cite{bek}, where the standard entropy combined with the BH horizon entropy is a raising function of time \cite{bek,uw} where, at current times, it is not always valid in the case of the alternative theories of GR although  it works when GR is considered.  

Recently, J. D. Barrow \cite{barrow} described a scenario where quantum gravity effects could carry some intricate, fractal structure on the BH surface and consequently, changing its real horizon area  \cite{geral}. This scenario headed us to a new BH entropy expression given by
\bee
S_{Barrow} = \Bigg(\frac{A}{4G} \Bigg)^{1+\Delta/2}\,\,,
\eee

\ni where $A$ is the usual horizon area and $4G$ is the Planck area. Hence, we believe that still there exist questions about the behavior of the GSL in the light of specific thermostatistical approaches.  In other words, some questions were not responded such like the ones about the conditions on the validity of the GSL when thermostatistical approaches are taken into consideration.   Since GSL holds for the standard entropy, but it is conditionally violated in the extended entropies, the consequence is that the parameter of these entropies should be constrained in small values in order for the GSL to be satisfied.  In  the search of such explanations, in this work we have investigated some relevant entropies in the light of the second law of thermodynamics and we verified the conditions of their applications.

In this work we organized the issues in the following way, in section 2 we described in a summary way, the statistical approaches used here.   In section 3, we described the GSL and we computed the range of GSL violation for each statistical application.   In section 4 we presented the numerical results for each statistical formalism described in section 2.   In section 5, the conclusions were depicted.



\section{The statistical approaches}

It is important to say that nongaussian entropies have become very interesting in the investigation of the thermodynamics of the universe. Among them we can mention 
Tsallis entropy nonextensive approach \cite{tsallis} which is defined by
\begin{eqnarray}
	\label{tsad}
	S_T = k_B \,\frac{1}{1-q} \, \sum_{i=1}^W \Big(p_i^q - p_i\Big) \,\,,
\end{eqnarray}

\ni where $W$ is the number of discrete configurations,  $p_i$ denotes the ordinary probability of accessing state $i$ and $q$ is the parameter that measures the degree of nonextensivity. The definition of entropy in Tsallis statistics has the standard properties of equiprobability, concavity, extensivity but not additivity. This approach has been successfully applied in many different physical systems. As examples, we can mention the Levy-type anomalous diffusion \cite{levy}, turbulence in a pure-electron plasma \cite{turb} and gravitational systems \cite{sys,sa,eu,maji,mora2}.
It is notable to mention that Tsallis thermostatistics formalism has the Boltzmann-Gibbs (BG) statistics as a particular case in the limit $q \rightarrow 1$ where the standard additivity of entropy can be recovered.

In addition, there is another important $q$-generalized entropy defined as
\begin{eqnarray}
	\label{renyi}
	S_R = k_B \,\frac{1}{1-q} \, \ln \sum_{i=1}^W \, p_i^q \,\,,
\end{eqnarray}

\ni which is known as R\'enyi entropy \cite{renyi}. Combining both Eqs. (\ref{renyi}) and (\ref{tsad}), and using the usual normalization
condition $\sum^W_{i=1} p_i=1$ we have that
\begin{eqnarray}
	\label{ret}
	S_R= \frac{k_B}{\lambda} \, \ln \Bigg(1+\lambda \frac{S_T}{k_B} \Bigg) \,\,,
\end{eqnarray}

\ni where  $\lambda \equiv 1-q$ is also a constant parameter. If we take the limit $\lambda \rightarrow 0$ in Eq.  (\ref{ret}) then we have that 
$S_R=S_T$, where $S_T$ is the Tsallis entropy as defined in Eq. (\ref{tsad}).
This modified R\'enyi entropy (MRE) model was suggested initially  by Bir\'o and Czinner \cite{ci}.  It was also used by other authors, for example in ref. \cite{ko,many}.   Bir\'o and Czinner considered Tsallis entropy, $S_T$, as the Bekenstein-Hawking (B-H, to differentiate from BH) entropy \cite{swh,jdb} and they wrote the MRE as a function of $S_{B-H}$, such that
\begin{eqnarray}
	\label{sar}
	S_R= \frac{k_B}{\lambda} \, \ln \Bigg(1+\lambda \frac{S_{B-H}}{k_B} \Bigg) \,\,.
\end{eqnarray}

On the other hand, the well known Kaniadakis statistics \cite{kani}, also known as a $\kappa-$ statistics, analogously to the Tsallis thermostatistics model, generalizes the usual BG entropy in the following form 
\begin{eqnarray}
	\label{kani}
	S_\kappa = -k_B \sum_i^W \frac{p_i^{1+\kappa} -  p_i^{1-\kappa}}{2\kappa}  \,,
\end{eqnarray}

\ni where in the limit $\kappa \rightarrow 0$ the BG entropy is recovered. It is important to mention here that the $\kappa$-entropy satisfies the usual entropy properties.   Among them we can cite, for example, equiprobability, concavity and extensivity. The $\kappa$-statistics, just like Tsallis' statistics, has been successful when applied in many experimental scenarios. For example, the cosmic rays \cite{kani2}, cosmic effects \cite{nos2} and gravitational systems \cite{nosk}. Using the microcanonical ensemble definition, where all the states have the same probability, the Kaniadakis entropy reduces to \cite{kani}
\begin{eqnarray}
	\label{kanim}
	S_\kappa = k_B \frac{W^\kappa-W^{-\kappa}}{2\kappa} \,,
\end{eqnarray}

\ni and in the limit $\kappa \rightarrow 0$, we recover the usual BG entropy formula, $S = k_B \ln  W$.

The authors of \cite{bc,ci} have proposed a novel type of R\'enyi entropy on BH horizons by considering the B-H entropy as a nonextensive Tsallis entropy and taking a logarithmic formula.    The result is a dual $\,$Tsallis entropy which can be obtained and, due to its nonextensive effects, the BH can be in a stable  thermal equilibrium.   Motivated by this result we will consider that the Kaniadakis entropy, Eq. (\ref{kanim}), describes the B-H entropy
\begin{eqnarray}
\label{kbh}
k_B \frac{W^\kappa-W^{-\kappa}}{2\kappa}= S_{B-H} \,.
\end{eqnarray}

\ni Solving Eq. (\ref{kbh}) we have
\begin{eqnarray}
\label{wk}
W=\left(\kappa\, \frac{S_{B-H}}{k_B} \pm \sqrt{1+\kappa^2\, \frac{S_{B-H}^2}{k_B^2}} \right)^\frac{1}{\kappa} \,.
\end{eqnarray}

\ni Now, to use Eq. (\ref{wk}) into BG entropy make us to choose the positive sign from Eq. \eqref{wk} since we will substitute $W$ within a logarithm relation.   So, 
we obtain
\begin{eqnarray}
\label{bgkd}
S^*_\kappa= k_B \ln\, W =  \frac{k_B}{\kappa}\,  \ln \left(\kappa\, \frac{S_{B-H}}{k_B} + \sqrt{1+\kappa^2\, \frac{S_{B-H}^2}{k_B^2}}\right) \,\,,
\end{eqnarray}

\ni and when we make $\kappa=0$ in Eq. (\ref{bgkd}), $S^*_\kappa \rightarrow S_{B-H}$.   The expression in Eq. (\ref{bgkd}) is a BG deformation of the Kaniadakis entropy and we will call $S^*_\kappa$ as the dual Kaniadakis entropy (DKE). So, we can say that DKE is the corresponding version of the MRE approach in 
Kaniadakis statistics. This proposal has been applied in BH thermodynamics \cite{nosdual}.


\section{The generalized second law of thermodynamics}

The  GSL of thermodynamics was proposed initially by Bekenstein \cite{bek}, and some applications of his suggestion  were made for example 
in wormhole geometry \cite{boka, rak}.  To make this work more clear we will make here a brief review about the procedure described in ref. \cite{sb}. From now 
on we will use that $\hbar=c=k_B=1$. Consider a Friedmann-Robertson-Walker metric given by
\begin{eqnarray}
	\label{FRW}
	ds^2= - dt^2 + a^2(t) \delta_{ij} dx^i dx^j   \,\,.
\end{eqnarray}

\ni Thus, both Friedmann equations are
\begin{eqnarray}
	\label{fried1}
	H^2&=&\frac{8 \pi G}{3} \left(\rho_m + \rho_{DE} \right)    \,\,,\\
    \label{fried2}
	\dot{H}&=&-4 \pi G \left(\rho_m + p_m + \rho_{DE} + p_{DE} \right)    \,\,,
\end{eqnarray}

\ni where $H$ is the Hubble parameter defined as $H \equiv\dot{a}/a$ and the dot denotes the time derivative.
From Eqs. (\ref{fried1}) and (\ref{fried2}) we can observe that the universe is considered to be filled with matter and dark energy,
perfect fluids with energy density and pressure denoted by ($\rho_{DE},\; \rho_m$) and ($p_{DE},\;p_m$) respectively. The conservation
of the total energy-momentum tensor yields
\begin{eqnarray}
	\label{cons}
	\dot{\rho}_{DE} + 3 H \left(1+\omega_{DE} \right) \rho_{DE} + \dot{\rho}_m + 3 H \left(1+\omega_m \right) \rho_m = 0 \,\,\,,
\end{eqnarray}

\ni where $\omega_{DE} = p_{DE}/\rho_{DE}$ means the equation of state parameter of the dark energy and  $\omega_m = p_m/\rho_m$ means the equation of state parameter of the matter. It is important to say that in this work we have just considered the standard case, which means that both sectors do not interact.
The universe horizon will be regarded as the apparent horizon where its radius is given by 
\begin{eqnarray}
	\label{horizon}
	   \tilde{r}_A = \frac{1}{\sqrt{H^2+\frac{k}{a^2}}} \,\,\,.                                    
\end{eqnarray}

\ni The constant $k$ defines the spatial curvature and we will use it equal to zero in this work which leads us to 
\begin{eqnarray}
	\label{horizon2}
	\tilde{r}_A = \frac{1}{H} \,.                                    
\end{eqnarray}

\ni The first law of thermodynamics applied to the universe with dark energy and matter can be written in a differential form as
\begin{eqnarray}
\label{dsd}
dS_{DE} &=& \frac{1}{T} \left(P_{DE} dV + dE_{DE}  \right) \,,\\
\label{dsm}
dS_m &=& \frac{1}{T} \left(P_m dV + dE_m  \right) \,.                            
\end{eqnarray}

\ni If we multiply both Eqs. (\ref{dsd}) and (\ref{dsm}) by $1/dt$, and using Eq. (\ref{cons}), we have the following relations 
\begin{eqnarray}
\label{dra}
\dot{\tilde{r}}_A &=& - \dot{H} \tilde{r}_A^2 \,,\\
\label{ede} 
E_{DE}&=&\frac{4\pi}{3} \tilde{r}_A^3 \,\rho_{DE} \,,\\
\label{edm}
E_m&=&\frac{4\pi}{3} \tilde{r}_A^3 \,\rho_m \,,
\end{eqnarray}

\ni and we obtain that
\begin{eqnarray}
	\label{dotsde}
	\dot{S}_{DE} &=& \frac{1}{T} (1+\omega_{DE} )\, \rho_{DE}\, 4\pi \tilde{r}^2_A\, (\dot{\tilde{r}}_A - H \tilde{r}_A ) \,\,\,,\\
        \mbox{} \nonumber \\
	\label{dotsm}
	\dot{S}_m &=&  \frac{1}{T} (1+\omega_m )\, \rho_m\, 4\pi \tilde{r}^2_A\, (\dot{\tilde{r}}_A - H \tilde{r}_A ) \,\,\,.                            
\end{eqnarray}

\ni Considering that the apparent horizon temperature has the same form of the BH horizon temperature then we have
\begin{eqnarray}
	\label{th}
	 T_h = \frac{1}{2 \pi \tilde{r}_A} = \frac{H}{2\pi}\,,                      
\end{eqnarray}

\ni where we have used expression in Eq. (\ref{horizon2}).    Hence, using Eqs. (\ref{fried1}), (\ref{fried2}), (\ref{horizon2}), (\ref{dra}), (\ref{dotsde}), 
(\ref{dotsm}) and (\ref{th})  we can write the time derivative of the entropy inside the apparent horizon as
\begin{eqnarray}
	\label{sdott1}
	\dot{S}_V &= &\dot{S}_{DE} + \dot{S}_m \nonumber \\
	&=& 8 \pi ^2 \tilde{r}^3_A (\dot{\tilde{r}}_A - H \tilde{r}) \Big[(1+ \omega_{DE})\rho_{DE}+(1+ \omega_m)\rho_m \Big] \nonumber\\
	&=& \frac{2\pi}{G} H^{-5} \dot{H} \left[\dot{H} + H^2 \right] \,\,\,.
\end{eqnarray}

In order to explain our procedure, firstly we consider the horizon entropy described by the B-H entropy  which can be written as $S_{B-H} = A/4G$, where $A$ is the horizon area \cite{swh,jdb}. So, the horizon entropy can be written in terms of the apparent horizon radius, $\tilde{r}_A$ and it can be given by
\begin{eqnarray}
	\label{sha}
	S_h = \frac{\pi\tilde{r}^2_A}{G} \,.
\end{eqnarray}

\ni since $A=4 \pi \tilde{r}^2_A$.   Differentiating Eq. (\ref{sha}) and using Eqs. (\ref{horizon2}) and (\ref{dra}) we have that
\begin{eqnarray}
	\label{shadot}
	\dot{S}_h = \frac{2 \pi \tilde{r}_A \dot{\tilde{r}}_A}{G} = - \frac{2 \pi}{G} H^{-3} \dot{H} \,.
\end{eqnarray}

\ni Therefore the time derivative of the total entropy is
\begin{eqnarray}
	\label{sdott2}
	\dot{S}_T&=& \dot{S}_V + \dot{S}_h \nonumber\\
	&=&\frac{2\pi}{G} H^{-5} \dot{H} \left[\dot{H} + H^2 \right] - \frac{2\pi}{G} H^{-3} \dot{H} \\
	\label{sdott3}
	&=& \frac{2\pi}{G} \, H^{-5} \dot{H}^2 \geq 0 \,\,,
\end{eqnarray}

\ni which is a non-decreasing function of time, and hence Eq. \eqref{sdott2} shows that the GSL of thermodynamics holds in a universe filled with matter and dark energy bounded by the apparent horizon whose entropy is described by the B-H area entropy law.   The case $\dot{H}=0$ which implies $\dot{S}_T=0$ corresponds to a de Sitter universe.

Here we will consider that the horizon entropy is described by the MRE in Eq. (\ref{sar}), which can be written as
\begin{eqnarray}
 	\label{sr}
 	S_h^{MRE} =\frac{1}{\lambda} \, \ln \Big(1+\lambda S_{B-H} \Big) 
 	=\frac{1}{\lambda} \, \ln \Bigg(1+ \frac{\lambda\, \pi \,\tilde{r}^2_A}{G}\Bigg) \,\,.
 \end{eqnarray}

\ni Differentiating Eq. (\ref{sr}) and using Eqs. (\ref{horizon2}) and (\ref{dra}) we can write that
\begin{eqnarray}
	\label{srdot1}
	\dot{S}_h^{MRE}  &=& \frac{2\pi \tilde{r}_A \dot{\tilde{r}}_a}{G}\Bigg(1+ \frac{\lambda \pi \tilde{r}_A^2}{G}\Bigg)^{-1}
	= -\frac{2 \pi}{G} \, \alpha \, H^{-3} \dot{H} \,,
\end{eqnarray}

\ni where 
\begin{eqnarray}
	\label{alfa}
	\alpha \equiv \frac{1}{1+\frac{\lambda\,\pi}{G H^2}} \,.
\end{eqnarray}

\ni Therefore the time derivative of the total entropy, $S_T^{MRE}$ in MRE approach is
\begin{eqnarray}
	\label{sdotr1}
	\dot{S}_T^{MRE} & = &\dot{S}_V + \dot{S}^{MRE}_h \nonumber\\
	&=& \frac{2\pi}{G} H^{-5} \dot{H} \left[\dot{H} + H^2 \right]
	-\frac{2 \pi}{G}\, \alpha\, H^{-3} \dot{H} \nonumber\\
    \label{srdot2}
	 &=& \frac{2\pi}{G} H^{-5} \dot{H} \, \left[ \dot{H} + (1-\alpha) \, H^2\right] \,.
\end{eqnarray}

\ni From Eq. (\ref{srdot2}) we can observe that if we make $\alpha=1$ ($\lambda=0$ in Eq. (\ref{alfa})) we recover the usual case $\frac{2\pi}{G} H^{-5} \dot{H}^2 \geq 0$ which corresponds to the BH area entropy law describing the apparent horizon, as expected. We can see that $\dot{S}^{MRE}_T$ is equal to zero if $\dot{H}=0$ which is a feature of the de Sitter universe. Here we will determine the range of the extra parameter $\lambda$ which in Eq. (\ref{srdot2}) will be negative and consequently leading to a GSL of thermodynamics to be violated. The first case that we will consider is  $\dot{H} < 0$ (which corresponds to the universe fluids satisfy the null energy condition). Then from Eq. (\ref{srdot2}) we have that $\dot{H} + (1-\alpha) H^2 > 0$. After a little algebra we have that
\begin{eqnarray}
	\label{alfai}
\dot{H} < 0 \;\;\; \Rightarrow \;\;\; \alpha < 1 + \frac{\dot{H}}{H^2} \,.
\end{eqnarray}

\ni The second case is $\dot{H} > 0$ and  $\dot{H} + (1-\alpha) H^2 < 0$. Therefore we have the condition 
\begin{eqnarray}
	\label{alfaii}
	\dot{H} >  0 \;\;\; \Rightarrow \;\;\; \alpha >  1 + \frac{\dot{H}}{H^2} \,.
\end{eqnarray}

\ni If Eq. (\ref{alfai}) or (\ref{alfaii}) is satisfied then the GSL of thermodynamics is not obeyed in the MRE approach.

Considering the Sharma-Mittal statistics (SME) \cite{sham,xxx} we have that
\bee
\label{A}
S^{SME}_h\,=\, \frac{1}{1-r} \Bigg[ \Big( 1\,+\,(1-q) S_{B-H} \Big)^{\frac{(1-r)}{(1-q)}} \,-\,1 \Bigg] \,\,,
\eee

\ni where $r$ is a new parameter and $q$ is the Tsallis nonextensive parameter. In the limit $r \rightarrow 1$, the Sharma-Mittall entropy becomes R\'enyi entropy while for $r \rightarrow q$ we have Tsallis entropy. In the limiting case where both parameters $r$ and $q$ become 1, we recover the usual BG entropy.
Using Eq. \eqref{sha} we can rewrite Eq. \eqref{A} as
\bee
\label{B}
S^{SME}_h\,=\,\frac 1R \Bigg[ \Bigg( 1\,+\, \frac{ \delta\,\pi\tilde{r}^2_A}{G} \Bigg)^Q\,-\,1\Bigg]\,\,,
\eee

\ni where $Q=R/\delta$, $R=1-r$ and $\delta=1-q$. Differentiating Eq. (\ref{B}) and using Eqs. (\ref{horizon2}) and (\ref{dra}), that are 
$\tilde{r}_A=H^{-1}$ and $\dot{\tilde{r}}_A=-\dot{H} \tilde{r}^2_A\, $ respectively, we have

\bee
\label{C}
\dot{S}^{SME}_h\,=\,-\,\frac{2\pi}{G}\,H^{-3}\,\dot{H} \Bigg( 1\,+\, \frac{\delta\pi}{G}\,H^{-2} \Bigg)^{Q-1} \,\,.
\eee

\ni Consequently, the time derivative of the total entropy, $S^{SME}_T$ in the SME scenario is
\bee
\label{C.1}
\dot{S}^{SME}_T\,=\,\dot{S}_V\,+\,\dot{S}^{SME}_h \,,
\eee

\ni and again, using Eq. \eqref{sdott1} and now the Eq. \eqref{C}, the final result of $\dot{S}^{SME}_T$ is
\bee
\label{D}
\dot{S}^{SME}_T\,=\,\frac{2\pi}{G}\,H^{-5}\dot{H} \Bigg\{\dot{H}\,+\,H^2\, \Bigg[ 1\,-\,\Bigg(1\,+\,\frac{\delta\pi}{G}\,H^{-2}\Bigg)^{Q-1}\,\,\Bigg]\Bigg\}\,\,.
\eee

\ni When we make $Q=1$ in (\ref{D}) we have $\dot{S}^{SME}_T\,=\,(2\pi/G)H^{-5}\dot{H}^2$, which is positive and consequently the GSL holds.
We can see that $\dot{S}^{SME}_T$ is equal to zero if $\dot{H}=0$ which is a de Sitter universe feature.  We will establish the range of the extra parameters $\delta$ and $Q$ where the GSL of thermodynamics will be violated. The first case that we will consider is  $\dot{H} < 0$. Then from Eq. (\ref{D}) we have that  $\dot{H} + H^2 [ 1 - (1+ \frac{\delta \pi}{G}\, H^{-2} )^{Q-1} ] > 0$. Consequently after an algebraic work we obtain
\begin{eqnarray}
	\label{Qi}
	\dot{H} < 0 \;\;\; \Rightarrow \;\;\; \left( 1 + \frac{\delta \pi}{G} H^{-2} \right)^{Q-1} < \, 1 + \frac{\dot{H}}{H^2} \,\,.
\end{eqnarray}

\ni The second case is $\dot{H} > 0$ and $\dot{H} + H^2 [ 1 - (1+ \frac{\delta \pi}{G}\, H^{-2} )^{Q-1} ] < 0$. Therefore we have
\begin{eqnarray}
	\label{Qii}
	\dot{H} > 0 \;\;\; \Rightarrow \;\;\; \left( 1 + \frac{\delta \pi}{G} H^{-2} \right)^{Q-1} > \, 1 + \frac{\dot{H}}{H^2} \,\,.
\end{eqnarray}

\ni Therefore the GSL of thermodynamics when the apparent horizon is described by the SME is no longer valid if the extra parameters $\delta$ and $Q$ have
the conditions Eq. (\ref{Qi}) or (\ref{Qii}).

In the DKE case the horizon entropy is given by
\begin{eqnarray}
	\label{shdka}
	S_h^{DKE} &=& \frac{1}{\kappa}\,  \ln \left(\kappa\, \frac{\pi \tilde{r}^2_A}{G} + \sqrt{1+\kappa^2\, \left(\frac{\pi \tilde{r}^2_A}{G}\right)^2}\right) \,.
\end{eqnarray}

\ni Differentiating Eq. (\ref{shdka}) and using Eqs. (\ref{horizon2}) and (\ref{dra}) and after some algebra we obtain
\begin{eqnarray}
	\label{shdotdka}
	\dot{S}^{DKE} _h = \frac{\frac{2\pi \tilde{r}_A \dot{\tilde{r}}_a}{G}}{\sqrt{1+\left(\frac{\kappa\pi \tilde{r}_A^2}{G}\right)^2} }
			=- \frac{2\pi}{G} \, \beta \, H^{-3} \dot{H}  \,,
\end{eqnarray}

\ni where 
\begin{eqnarray}
	\label{beta}
\beta \equiv \frac{1}{\sqrt{1+\left(\frac{\kappa\pi}{G H^2}\right)^2}}\,. 
\end{eqnarray}

\ni From Eq. (\ref{beta}) we can see that the range of $\beta$ parameter is 
\begin{eqnarray}
\label{range}
 0 < \beta \leq 1 \,.
\end{eqnarray}	

\ni The time derivative of the total entropy, $S_T^{DKE}$,  using again Eq. \eqref{sdott1}, is given by
\begin{eqnarray}
	\label{stdotdka}
	\dot{S}^{DKE} _T &=& \dot{S}_V + \dot{S}^{DKE}_h \nonumber\\
	                 &=& \frac{2\pi}{G} \, H^{-5} \dot{H} \left[ \dot{H}+H^2 \right]- \frac{2\pi}{G} \, \beta \, H^{-3} \dot{H} \nonumber\\
	\label{stdotdka2}                 
	                 &=& \frac{2\pi}{G} H^{-5} \dot{H}\, \left[ \dot{H} + \left(1 - \beta \right) H^2\right] \,.
\end{eqnarray}

\ni From Eq. (\ref{stdotdka}) we can observe that we assume $\beta=1 $, $\kappa=0$ in Eq. (\ref{beta}), then $\dot{S}^{DKE} _T$ becomes $\frac{2\pi}{G} H^{-5} \dot{H}^2 \geq 0$ which corresponds to the BH area entropy law describing the apparent horizon. We can observe that if we make $\dot{H}=0$ in (\ref{stdotdka}) then $\dot{S}^{DKE} _T$ is equal to zero and, as we have already mentioned in the MRE and SME cases, this is feature of the de Sitter universe.  We will establish the range of the extra parameter $\beta$ where the GSL of thermodynamics will be violated. The first case that we will consider is  $\dot{H} < 0$. Then from Eq. (\ref{stdotdka}) we have that  $\left[ \dot{H} + \left(1 - \beta \right) H^2\right] > 0 $. Consequently after a little algebra we have
	\begin{eqnarray}
		\label{bi}
		\dot{H} < 0 \;\;\; \Rightarrow \;\;\; \beta < 1 + \frac{\dot{H}}{H^2} \,.
	\end{eqnarray}

\ni So, due to the range of $\beta$ parameter, Eq. (\ref{range}), we have only one condition, Eq. (\ref {bi}), where the GSL of thermodynamics when the apparent horizon is described by the DKE is no longer be valid.


\section{Numerical results}

 In order to investigate the evolution of $\dot{S}_T$ we will consider the Hubble parameter as a function of time. For convenience we use the redshift, defined as $1+z=a_0/a$, where $a_0=1$ the present scale factor, as the independent variable. We will assume that the way which the Hubble function evolves will be given by the $\Lambda$CDM cosmology defined as
\begin{eqnarray}
	\label{cdm}
	H_{\Lambda CDM}(z) = H_0 \sqrt{ \Omega_{m0}(1+z)^3+\Omega_{r0}(1+z)^4+\Omega_{\Lambda0}} \,\,,
\end{eqnarray}

\ni where $H_0$ is the current Hubble parameter, $\Omega_{m0}$ is the value of the matter density parameter, $\Omega_m = 8 \pi G \rho_m/(3H^2)$, $\Omega_{r0}$ is the value of the radiation density parameter, $\Omega_r = 8 \pi G \rho_r/(3H^2)$ and $\Omega_{\Lambda0} = 1-\Omega_{m0} - \Omega_{r0}$.  The 0-index means the current value of the parameter.

Firstly, let us consider the MRE approach. So, we can write Eq. (\ref{srdot2}) as a function of the redshift z as
\begin{eqnarray}
	\label{mrez}
	\dot{S}_T^{MRE}(z) = \frac{2\pi}{G} (1+z) \, H(z)^{-3} H'(z)\, \left[ (1+z)\, H'(z) - (1-\alpha) \, H(z) \right] \,,
\end{eqnarray}

\ni where primes means derivatives with respect to $z$ and we have used the relation $\dot{H} = - (1+z) H H'$. Consequently, using the $\Lambda$CDM cosmology, Eq. (\ref{cdm}),
we have
\begin{eqnarray}
	\label{cdml}
	H'_{\Lambda CDM}(z) = \frac{H_0}{2} \frac{3 \Omega_{m0}(1+z)^2+4 \Omega_{r0}(1+z)^3}{\sqrt{\Omega_{m0}(1+z)^3+\Omega_{r0}(1+z)^4+\Omega_{\Lambda0}}}\,.
\end{eqnarray}

\ni Using Eqs. (\ref{cdm}) and (\ref{cdml}) we have plotted in Fig. 1 the evolution of $\dot{S}^{MRE}_T(z)$, Eq. (\ref{mrez}), for several values of the $\alpha$ parameter. We have chosen $\Omega_{m0}=0.30$,
$\omega_{r0}=0.0001$ and $G=H_0=1$ \cite{sb}. We can see that the GSL of thermodynamics is not respected in the MRE approach for the $\alpha$ 
parameter less than one and for small values of $z$.
\begin{figure}[H]
	\centering
	\includegraphics[height=6.cm,width=7.cm]{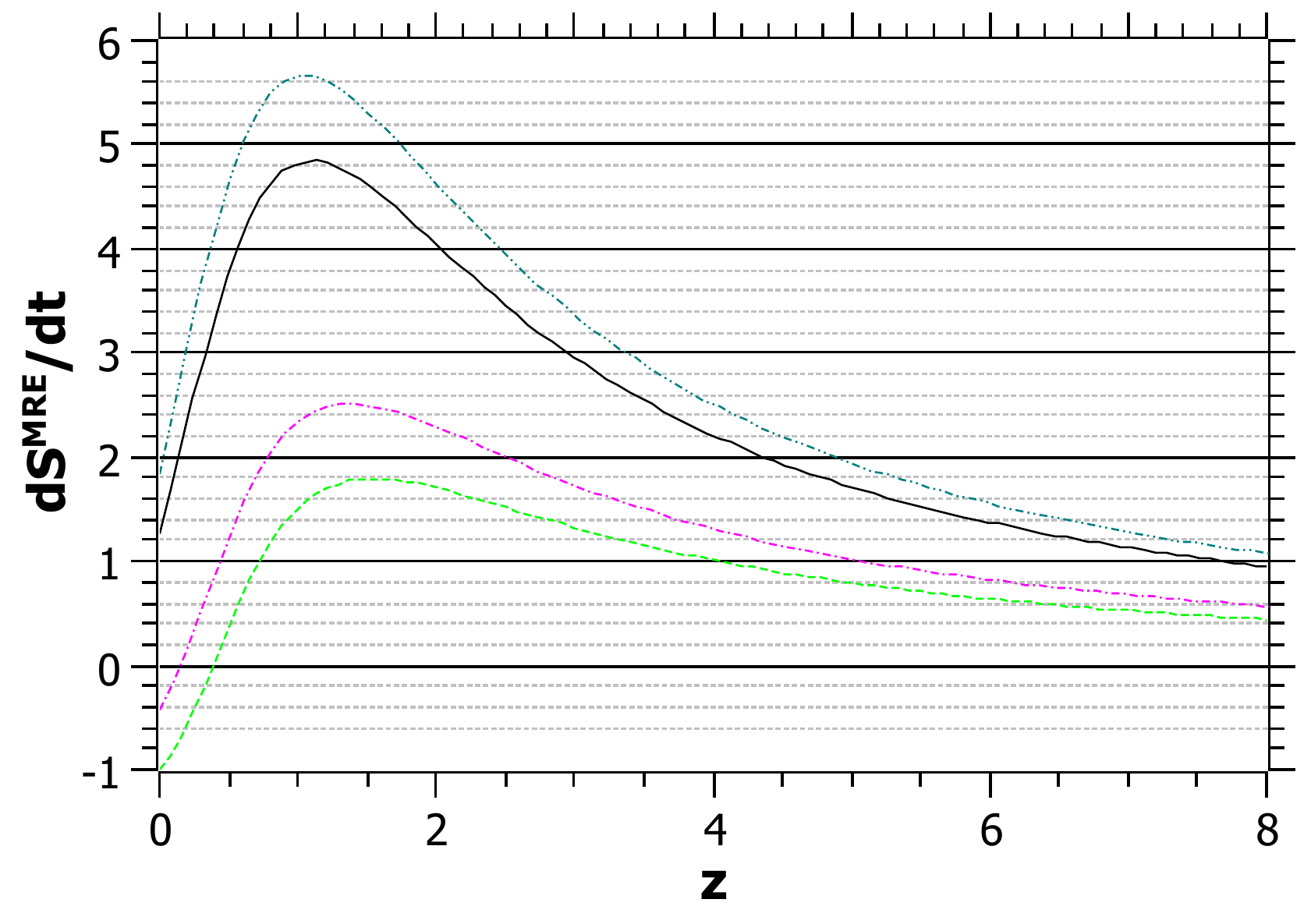}
	\caption{ $dS^{MRE}(z)/dt \equiv \dot{S}^{MRE}_T(z)$ versus the redshift $z$ for various values of the $\alpha$ parameter: $\alpha=1$ (black - line),  
	$\alpha=1.2$ (blue - dashed-dotted-dotted), $\alpha=0.4$ (red - dashed-dotted) and $\alpha=0.2$ (green - dashed).}
	\label{dmre}
\end{figure}

The second case is the SME model. In a similar way to the MRE approach, we can write Eq. (\ref{D}) as a function of the redshift z as
\begin{eqnarray}
	\label{smez}
	&&\dot{S}_T^{SME}(z) \\
&=& \frac{2\pi}{G} (1+z) \, H(z)^{-3} \, H'(z)  \left\{ (1+z)\,H'(z) - \left[ 1-\left( 1 + \frac{\delta \pi}{G} H(z)^{-2}\right )^{Q-1} \right] H(z) \right\}\,\,.
\nonumber 
\end{eqnarray}

\ni Using the $\Lambda$CDM cosmology,  Eq. (\ref{cdm}), we have plotted the evolution of $\dot{S}^{SME}_T(z)$ for several values of $Q$ in Fig. 2. Again we have chosen $\Omega_{m0}=0.30$, $\omega_{r0}=0.0001$ and $H_0=G=\delta=1$. We can see that the GSL of thermodynamics is not satisfied in the SME approach, i.e., Eq. (\ref{smez}) is negative if the $Q$ parameter is negative and for small values of $z$. 
\begin{figure}[H]
	\centering
	\includegraphics[height=6.cm,width=7.cm]{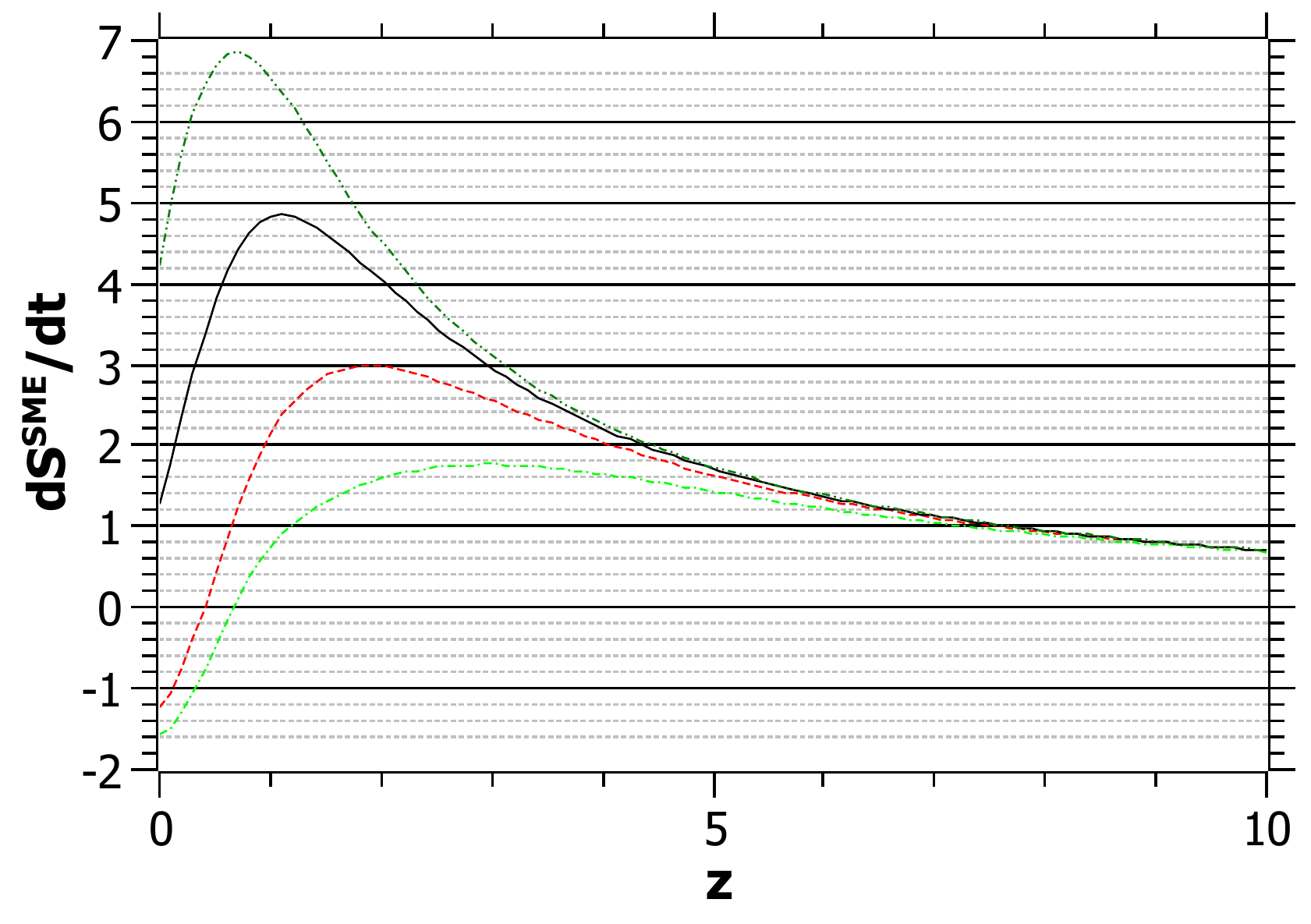}
	\caption{ $dS^{SME}(z)/dt \equiv \dot{S}^{SME}_T(z)$ versus the redshift $z$ for various values of the $Q$ parameter: $Q=1$ (black - line), $Q=1.5$ (blue - dashed - dotted - dotted),
	 $Q=-0.5$ (red - dashed) and $Q=-5$ (green - dashed - dotted).}
	\label{dsme}
\end{figure}

Finally, let us analyze the DKE model. In a similar way to the MRE and the SME approaches, we can write Eq. (\ref{stdotdka2}) as a function of the redshift z as
\begin{eqnarray}
	\label{dkez}
	\dot{S}_T^{DKE}(z) = \frac{2\pi}{G} (1+z) \, H(z)^{-3} H'(z)\, \Big[ (1+z)\, H'(z) - (1-\beta) \, H(z) \Big] \,.
\end{eqnarray}

\ni Using the $\Lambda$CDM cosmology from Eq. (\ref{cdm}), we have plotted the evolution of $\dot{S}^{DKE}_T(z)$, Eq. (\ref{dkez}), for several values of the $\beta$ parameter in Fig. 3. Once more we have chosen $\Omega_{m0}=0.30$, $\omega_{r0}=0.0001$ and $G=H_0=1$. In the DKE the $\beta$ parameter is less than or equal to one, Eq. (\ref{range}). So, as we can see in Fig. 3, the GSL of thermodynamics is not satisfied in DKE approach, i.e., $\dot{S}_T^{DKE}$ is negative for small values of $z$. Here it is important to comment that the behavior of $\dot{S}_T^{DKE}$, Eq. (\ref{dkez}), is similar to $\dot{S}_T^{MRE}$, Eq. (\ref{mrez}).

\begin{figure}[H]
	\centering
	\includegraphics[height=6.cm,width=7.cm]{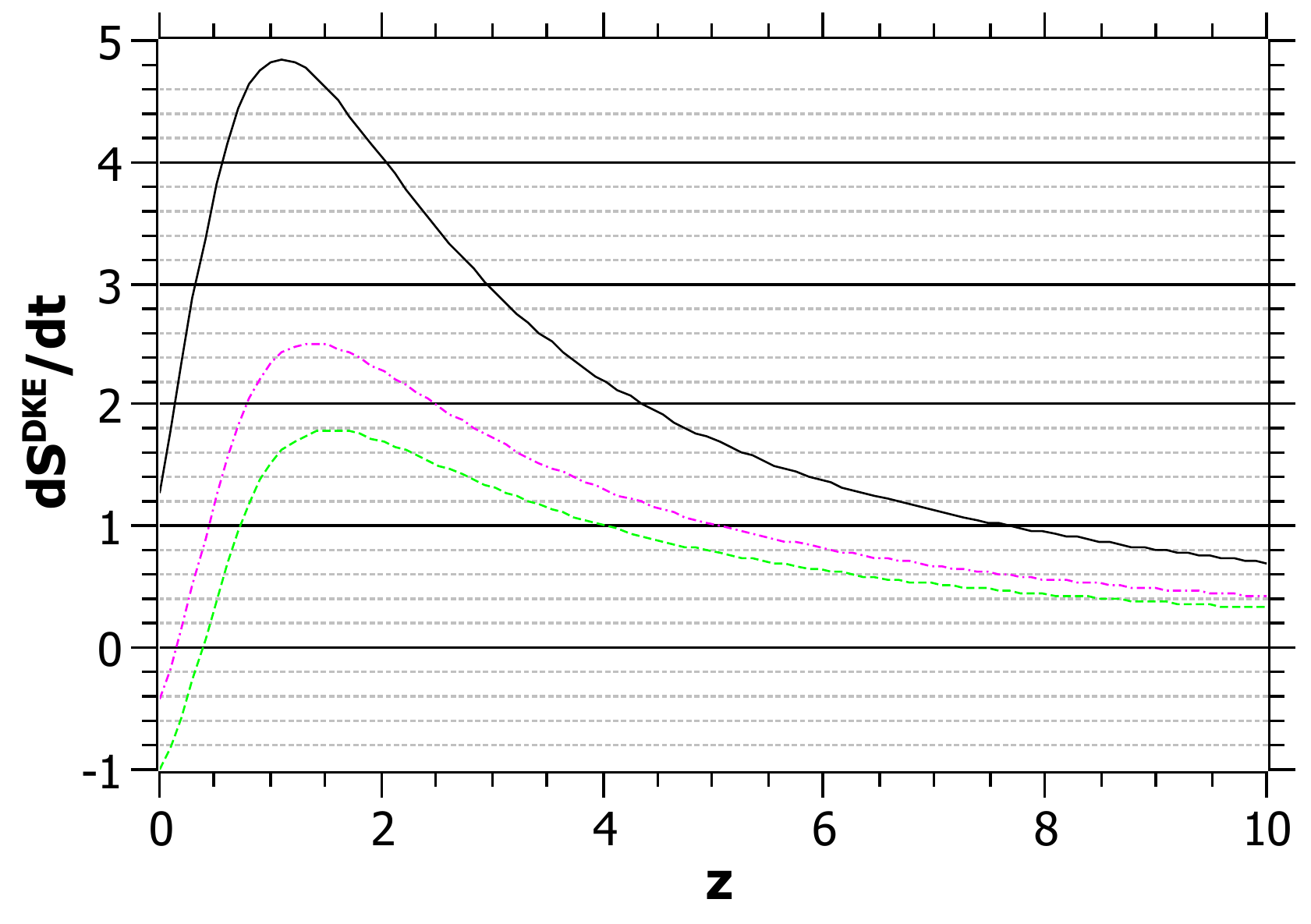}
	\caption{ $dS^{DKE}(z)/dt \equiv \dot{S}^{DKE}_T(z)$ versus the redshift $z$ for various values of the $\beta$ parameter: $\beta=1$ (black - line),  
	$\beta=0.4$ (red - dashed-dotted) and $\beta=0.2$ (green - dashed).}
	\label{dkeg}
\end{figure}



\section{Conclusions}

To conclude, in this work we have investigated the effect of MRE, SME and DKE approaches when we describe the apparent horizon which bounds a universe filled with dark energy and usual matter. Initially, considering that the universe does not evolve, we have obtained that for some special parameters the GSL of thermodynamics is not respected in MRE, SME and DKE. Considering the universe evolution when the $\Lambda CDM$ model was used, Eq. (\ref{cdm}), the result is that the GSL of thermodynamics for some specific conditions is not satisfied in MRE, SME and DKE. It is important to say that we have assumed that the usual Einstein general relativity continues to control the evolution of the universe. Only nongaussian entropies, which are extensions of BG entropy, were considered. As a perspective for future work, it would be interesting to investigate the effects of other important nongaussian entropies as well as alternative modified gravity theories from the point of view the GSL of thermodynamics.

\newpage

\section*{Acknowledgments}

\ni  We would like to thank the anonymous referee for useful comments. The authors also thank CNPq (Conselho Nacional de Desenvolvimento Cient\' ifico e Tecnol\'ogico), Brazilian scientific support federal agency, for partial financial support, Grants numbers  406894/2018-3 (Everton M. C. Abreu) and 307153/2020-7 (Jorge Ananias Neto).

\end{document}